\documentclass[twocolumn,showpacs,preprintnumbers,amsmath,amssymb]{revtex4}

\usepackage{graphicx}%
\begin{document}

\newcommand{\bc}{\begin{center}} \newcommand{\ec}{\end{center}}
\newcommand{\be}{\begin{equation}} \newcommand{\ee}{\end{equation}}
\newcommand{\beqn}{\begin{eqnarray}} \newcommand{\eeqn}{\end{eqnarray}}

\title{Partially asymmetric exclusion models with quenched disorder}

\author{R\'obert Juh\'asz, Ludger Santen}
 \email{juhasz,santen@lusi.uni-sb.de} \affiliation{
 Theoretische Physik, Universit\"at des Saarlandes, D-66041
 Saarbr\"ucken, Germany  }%

\author{Ferenc Igl\'oi}  \email{igloi@szfki.hu}
\affiliation{ Research Institute for Solid
State Physics and Optics, H-1525 Budapest, P.O.Box 49, Hungary}
\affiliation{ Institute of Theoretical Physics, Szeged University,
H-6720 Szeged, Hungary}

\date{\today}

\begin{abstract}
We consider the one-dimensional partially asymmetric exclusion process
with random hopping rates, in which a fraction of particles (or sites)
have a preferential jumping direction against the global drift. In
this case the accumulated distance traveled by the particles, $x$, scales
with the time, $t$, as $x \sim t^{1/z}$, with a dynamical exponent $z
> 0$. Using extreme value statistics and an asymptotically exact
strong disorder renormalization group method we exactly calculate, $z_{pr}$,
for particlewise (pt) disorder, which is argued to be related as, $z_{st}=z_{pr}/2$,
for sitewise (st) disorder. In the symmetric case
 with zero mean drift the particle diffusion is ultra-slow,
logarithmic in time.
\end{abstract}

\pacs{Valid PACS appear here}

\maketitle

Driven lattice gas models are able to describe the generic features of
stochastic transport far from equilibrium
\cite{hinrichsen,schutzreview,privman}.
 This kind of transport is
often observed, e.g. in transport systems \cite{chowd} or
(micro-)biological applications \cite{juelrmp}, where a
steady flow is maintained by a steady input of energy. As an important recent application
we mention the
bi-directional transport on microtubules \cite{gross}, in which
motion of cargos is realized by attaching a few molecular
motors of different type. In our approach a cargo-motor complex is
modeled as an effective particle with random hopping rates, the value of
which depends on the number of attached motors.
This kind of molecular machine allows
for an efficient navigation in the cell, because the velocity can
be controlled via the  typical number of attached motors.
Moreover these models
can be related via simple mappings to  other problems of
nonequilibrium statistical physics which include  e.g. surface growth
problems \cite{hinrichsen,schutzreview}.

Here we discuss the most prominent model of this kind the so-called
asymmetric simple exclusion process (ASEP), what is considered on
a periodic chain of $N$ sites and with $M$ particles. For particlewise (pt)
disorder particle $i$ may hop to empty neighboring sites with rates
$p_i$ to the right and $q_i$ to the left, where $p_i$ and $q_i$ are independent
and identically distributed random variables. For sitewise (st) disorder
the random hopping rates are assigned to given sites of the lattice. 
In the {\it totally asymmetric model} $q_i=0$, for every $i$, whereas in the {\it partially
asymmetric model}, $q_i>p_i$ for a finite fraction of $i$.

A characteristic feature of the ASEP and its variants is their
sensitivity to spatial inhomogeneities or quenched disorder of any
kind. This  is expressed in boundary induced phase transitions
\cite{krug} or phase separated states caused by a single defect
\cite{janowsky} or disordered lattices \cite{barma}. For pt disorder some exact
results are available \cite{kf,evans,evanspar}, in particular for the totally asymmetric
case. Depending on the extremal properties of the hopping rate
distribution there is a dynamical phase transition in the system,
separating a homogeneous state from a nonhomogeneous one, in which
there is a macroscopic particle free region in front of the domain of
occupied sites. For st disorder the
analytical results are scarce \cite{stanal} and most of our knowledge is based on Monte Carlo
simulations \cite{barma,derrida} and mean-field calculations \cite{barma,stinchcombe}.
In the totally asymmetric model as the strength of disorder is increasing in the
stationary state the particle density is changing from a homogeneous phase into
a segregated-density phase, in which macroscopic regions with densities
$0<\rho_c<1-\rho_c<1$ coexist and the macroscopic current, $J$, generally decreasing
with decreasing $\rho$. In the partially asymmetric model $\rho_c$, and at the
same time $J$ approaches zero in the thermodynamic limit. More precisely the
stationary velocity $v$ vanishes in a large ring of $N$ sites as:
\be v \sim N^{-z}\;,
\label{z_st}
\ee
and the accumulated distance traveled by the particles, $x$, in time, $t$, is given by
$x \sim t^{1/z}$, where $z$ is the dynamical exponent.

In this letter we consider the partially asymmetric disordered model and will show that a
vanishing current state can be encountered for pt disorder, too. We will derive an
analytical expression for $z_{pt}$ (pt disorder) and conjecture a simple relation
for $z_{st}$ (st disorder). We will also consider a new stationary state in which the average
drift is zero and the system has a diffusive motion. We show that this new
state for both types of disorder has an infinite randomness fixed
point (IRFP) scenario. So far IRFP has only been observed for quantum\cite{DF}
and stochastic models \cite{RGsinai,hiv} on disordered lattices.

Here we consider first the ASEP with pt disorder and a configuration
is characterized in terms of the number of empty sites, $n_i$, in front of the $i$th particle. The
stationary weight of a configuration $n_1,n_2,\dots ,n_M$ is given by
\cite{spitzer,evans}:  $f_N(n_1,n_2,\dots,n_M)=\prod_{\mu
=1}^Mg_{\mu}^{n_{\mu}}$, where  \be
g_{\mu}=\left[1-\prod_{k=1}^M{q_k\over
p_k}\right]^{-1}\left[\sum_{i=0}^{M-1}{1\over p_{\mu -i}}\prod_{j=\mu
+1-i}^{\mu}{q_j\over p_j}\right]
\label{g}
\ee provided $p_i> 0$ for all particles. The stationary
velocity is given by:
\be v=\frac{Z_{N-1,M}}{Z_{N,M}},\quad Z_{N,M}=\sum_{n_1,n_2,\dots,n_M} f_N(\{n_{\mu}\})\;.
\label{v}
\ee
where in the summation $\sum_{\mu=1}^M n_{\mu}=N-M$.
We are interested in the properties of the state in the thermodynamic
limit, where we define the control-parameter as:
\be \delta=\frac{[\ln p]_{\rm av} - [\ln q]_{\rm av}}{{\rm var}[\ln
p]+{\rm var}[\ln q]}\;,
\label{delta}
\ee
such that for $\delta>0$ ($\delta<0$) the particles move to the right
(left).  Here and in the following we use $[\dots]_{\rm av}$ to denote
averaging over quenched disorder and ${\rm var}[x]$ stands for the
variance of $x$. In the following we restrict ourselves to the domain $\delta \ge 0$.

The consequences of these formulae for random hopping rates have been
analyzed thoroughly if $v$ is non-vanishing in the thermodynamic
limit \cite{kf,evans,evanspar}, in particular for the totally asymmetric model.
For the partially asymmetric model we consider the maximal term, $g_{max}=\max\{g_{\mu}\}=g_{\mu^*}$,
which in Eq.(\ref{g}) is dominated by a product, $\prod_{a<j<b} q_j/p_j$, where $a<j<b$ is the
largest region in which $q_j>p_j$. The probability of existence of this rare region is,
$P(b-a) \sim \exp(-\alpha(b-a))$, therefore among $M$ particles its typical size is
$b-a \sim \ln M$. So we obtain: $g_{max} \sim \exp(\sigma(b-a)) \sim M^{\gamma}$, $\gamma>0$.
As a consequence, in the thermodynamic limit $Z_{N,M}$ is dominated
by that therm, in which $n_{\mu^*}$ is macroscopic and therefore the stationary velocity in Eq.~(\ref{v})
is given by: $v=g_{max}^{-1}$.

The distribution of the quantities, $g_{\mu}$, for a fixed $\mu$ and in the thermodynamic
limit $g_{\mu}=\lambda$ is, up to a prefactor, in the form of a
Kesten-variable \cite{kesten}. 
For large $\lambda$ this distribution takes the form,
$ P(\lambda)\sim \lambda^{-(1+1/z)}$,
where $z$ is the positive root of the equation
\be \left[\left({p/q}\right)^{1/z}\right]_{av}=1\;.
\label{z}
\ee
We argue that $z_{pt}=z$. For a large, but finite $M$,
and for a given realization of disorder two variables, $g_{\mu}$ and
$g_{\mu'}$, have negligible correlations, provided $|\mu-\mu'|>\xi$, where
$\xi$ is proportional to the (finite) correlation length
in the system. Consequently, the distribution of the $g_{\mu}$
variables in a given sample has the same power-law asymptotics with the exponent in
Eq.(\ref{z}). Since the stationary velocity is the
inverse of the largest $g_{\mu}$, the distribution of $v$ for different
samples is obtained from the statistics of extremes \cite{galambos}. Here we use the result
that the distribution of the maximum of independent random variables, which
are taken from a distribution with a power-law tail is
universal and given by the
Fr\'echet distribution: $P(u) = u^{-1-1/z}/z e^{-u^{-1/z}}$
with $u=g_{max}C M^{-z}$.
Thus $v \sim M^{-z}$ and with $M/N=O(1)$ we obtain from
Eq.~(\ref{z_st}) the announced result. For small $\delta$ the dynamical exponent
is divergent: $z \simeq (2\delta)^{-1}$.

\begin{figure}
\includegraphics[width=0.54\linewidth]{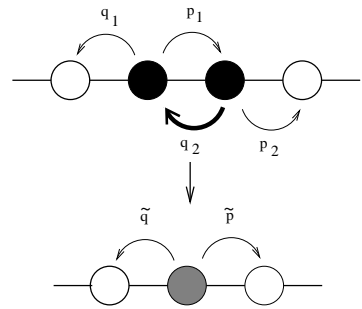}
\caption{\label{fig:renorm} Renormalization scheme for particle
  clusters. If $q_2$ is the largest hopping rate, in a time-scale, $\tau > 1/q_2$,
  the two-particle cluster moves coherently and the composite particle
  is characterized by the effective hopping rates $\tilde q$ and  $\tilde
  p$, respectively, see the text.}
\end{figure}

At $\delta=0$, when the stationary velocity in the system is vanishing and
the above formalism does not work, we use a strong disorder
renormalization group (RG) approach, which is analogous to that
applied recently for absorbing state phase transitions with quenched
disorder \cite{hiv} and originates in the theory of random quantum
spin chains \cite{MDH,DF} and random walks \cite{RGsinai}.
In the RG method one sorts the transition rates in descending order
and the largest one sets the energy scale,
$\Omega=max(\{p_i\},\{q_i\})$, which is related to the relevant
time-scale, $\tau=\Omega^{-1}$. During renormalization the largest
hopping rates are successively eliminated, thus the time-scale is
increased. In a sufficiently large time-scale some cluster of
particles moves coherently and form composite particles, which have
new effective transition rates. To illustrate the method (see Fig. \ref{fig:renorm})
let us assume
that the largest rate is associated to a left jump, say $\Omega=q_2$,
furthermore $q_2 \gg p_2,q_1,p_1$. In a time-scale, $\tau>\Omega^{-1}$, the
fastest jump with rate $q_2$ can not be observed and the two particles
$1$ and $2$ form a composite particle. The composite particle has a
left hopping rate $\tilde{q}=q_1$, since a jump of particle $1$ is
almost immediately followed by a jump of particle $2$. The transition
rate to the right, $\tilde{p}$, follows from
the observation that, if the neighboring site to the right of particle
$2$ is empty it spends a small fraction of time:
$r=p_2/(p_2+q_2)\approx p_2/q_2$ on it. A jump of particle $1$ to the
right is possible only this period, thus $\tilde{p}=p_1 r \approx p_1
p_2/q_2$. The renormalization rules can be obtained similarly for a
large $p$:
\be \tilde{p}=\frac{p_1 p_2}{\Omega},\quad \Omega=q_2;\quad
\tilde{q}=\frac{q_1 q_2}{\Omega} ,\quad \Omega=p_1\;.
\label{deci}
\ee
The RG scheme outlined above is completely equivalent to
that of a random antiferromagnetic (dimerized) $XX$ spin chain of $2M$
sites defined by the Hamiltonian:
\be H_{XX}=-\sum_{i=1}^{2M} J_i \left(S^x_{i}S^x_{i+1}+S^y_{i}S^y_{i+1}\right)\;,
\label{HXX}
\ee
with $J_{2i-1}=p_i$ and $J_{2i}=q_i$. Here $\delta$ in Eq.~(\ref{delta}) plays the role of the
dimerization. The strong disorder RG for the random XX-chain is
analytically solved \cite{fisherxx,ijr00} and the presumably asymptotically exact results\cite{note}
can be directly applied for the ASEP with pt disorder. In an extended part of the
off-critical regime, $\delta > 0$, the correlation length is finite, but the
typical time-scale, $t_r$, is divergent. This is the so-called
Griffiths phase \cite{griffiths} in which several dynamical quantities, such as
the susceptibility are singular. In the RG procedure in this phase almost
exclusively the left hopping rates are  decimated out
\cite{i02}. After $M$-steps of decimation we are left with a single
particle having a vanishing $\tilde{q}/\tilde{p}$ and $\tilde{p} \sim
M^{-z}$, with the dynamical exponent given in Eq.~(\ref{z}). This
completely coincides with our previous results.  At the critical
point, $\delta=0$, left and right hopping rates are decimated
symmetrically and the system scales into an IRFP. After $M$-steps the
remaining effective particle has a symmetric hopping probability:
$\tilde{q} \sim \tilde{p} \sim \exp(- const  M^{1/2})$. Thus the motion
of the system is diffusive and ultra-slow, the appropriate scaling
combination is given by: $\ln v M^{-1/2}$. Close to the critical point
the correlation length in the system, $\xi$, which measures the width
of the front, is given by $\xi \sim \delta^{-2}$.

\begin{figure}
\includegraphics[width=1.0\linewidth]{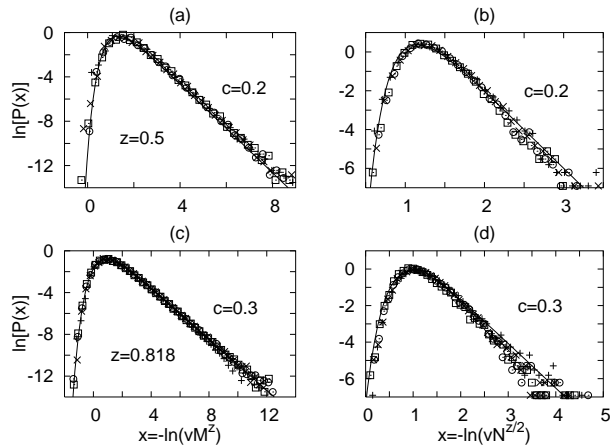}
\caption{\label{fig:griffiths} Scaling of the velocity distribution in the
Griffiths phase at two concentrations ($c=0.2$ and $c=0.3$) for pt
(a,c) and st (b,d) disorder.
Data denoted by symbols +,$\times$,O,$\Box$ correspond to
 $M=64,128,256,512$ for (a,c) and $N=64,128,256,512$ for (b,d), respectively.
The Fr\'echet distribution with the dynamical exponent: $z=z_{pt}$, calculated from Eq.(\ref{z}) and
with $z_{st}=z_{pt}/2$,
as given in Eq.(\ref{z_rel}) are indicated by a full line.}
\end{figure}

\begin{figure}
\includegraphics[width=0.8\linewidth]{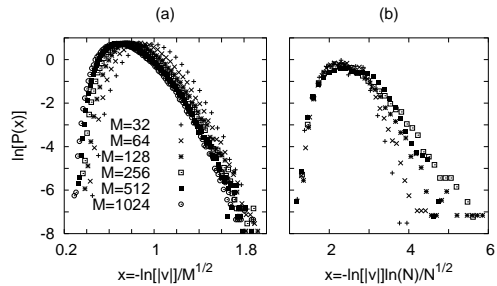}
\caption{\label{fig:critical} Scaling plot of the velocity
  distribution at the critical point for pt (a)
  and st (b) disorder. In the latter case we have included
  a logarithmic correction term.}
\end{figure}

These analytical results have been checked by calculating the velocity
distribution of a periodic system with $M/N=1/2$, using a
bimodal distribution with $p_iq_i=r$, for all $i$, and $P(p)=c
\delta(p-1)+(1-c) \delta(p-r)$, with $r>1$ and $0<c \le 1/2$. In
this case the control-parameter is $\delta=(1-2c)/[2c(1-c)
\ln r]$ and the dynamical exponent from Eq.~(\ref{z}) is $z=\ln
r/\ln(c^{-1}-1)$. (In the limit $c \ll 1/2$ one can make a direct
calculation by noting that an
extremely small velocity of $v(l)=C r^{-l}$ can be found in such a
sample, in which $l$ consecutive $q$ rates have the value of
$r>1$. Such a sample (rare event) is realized with an exponentially
small probability of $P(l)\simeq c^{l}$. Averaging over $l$ we arrive
to $z\simeq -\ln r / \ln c$.) As seen
in Fig.~\ref{fig:griffiths}ac in the Griffiths
phase for two different values of the
concentration, $c<1/2$, the distributions are well described by the
Fr\'echet statistics and the measured $z$ agrees very well with the
analytical result. At the critical point, $c=1/2$, as shown in Fig.~\ref{fig:critical}a,
a scaling collapse is obtained in terms of the scaling variable, $\ln
v M^{-1/2}$, which corresponds to the behavior at an IRFP.

For st disorder the rate of a jump to $i \to i+1$ ($i+1 \to
i$) is $p_i$ ($q_i$) and we use the same distribution, as
introduced before. In the small $c$ limit $z_{st}$ can
be calculated along the lines presented for pt disorder, but now
the velocity in a rare event of length $l$ is
\cite{bece00} $v(l) \simeq A r^{-l/2+1/4}$. The factor
$1/2$ in the exponential is due to the fact, that the boundary between
the $\rho=1$ and $\rho=0$ phases lies in the middle of the unfavorable domain
(particle-hole symmetry). For small concentration we obtain:
\be z_{st}=z_{pt}/2\;.
\label{z_rel}
\ee
According to our numerical
investigations, as shown in Fig.~\ref{fig:griffiths}bd, the relation in Eq.~(\ref{z_rel})
seems to hold for not small values of $c$, too.  In particular, at the critical situation, $c=1/2$,
we recover the result of the IRFP scenario, see Fig.~\ref{fig:critical}b.

In the following we present arguments, why the relation in
Eq.~(\ref{z_rel}) can be generally valid. The stationary state both for pt
and st disorder has a macroscopic phase
separation. However, for pt disorder the occupied region slowly moves,
which is due to single hole diffusion into the opposite direction, which
takes place in a position dependent random potential, the hopping rates of which,
$p_i$ and $q_i$, are generated by the particles. The stationary velocity is
the same as for a Sinai walker \cite{sinai}, see Eq.~(\ref{z}) and the relevant
time-scale in the problem, $\tau \sim v^{-1}$, is given by the time needed
for a single hole to overcome the largest barrier (rare event)
in the sample. This occurs with a probability of $p_{pt}(\tau) \sim \tau^{-1/z_{pt}}$,
since the typical value of $\tau=\tau_{pt}$ in a large sample
is given by $p_{pt}(\tau_{pt})N=1$, thus $\tau_{pt} \sim N^{z_{pt}}$, in
accordance with Eq.(\ref{z_st}).

On the other hand for the st disorder the position of the occupied block is fixed and
the diffusion of holes through the occupied phase and diffusion of
particles through the empty phase will result in the stationary
current. In this case, due to particle-hole symmetry, the rare event consists
of two (independent) large barriers, one for the holes and one for the
particles, both having the same time-scale. The probability of occurance of this rare event
is $p_{st}(\tau) \simeq p_{pt}^2(\tau)$.
Now the typical value of $\tau=\tau_{st}$ is given by $p_{st}(\tau_{st})N=1$,
thus $\tau_{st} \sim N^{z_{pt}/2}\sim N^{z_{st}}$, from which Eq.~(\ref{z_rel})
follows \cite{note2}.

In summary the ASEP with sufficiently strong particle or lattice
disorder has similar behavior. At $\delta=0$
in both cases the IRFP scenario holds, whereas for $\delta > 0$, in
the Griffiths phase we have a singular dynamical behavior governed by
a dynamical exponent, which is given in
Eqs.~(\ref{z}) and (\ref{z_rel}). The
case of pt disorder can be described through RG transformation as an
effective single-particle problem. For the st disorder case many particle
effects seem to be important. Our findings might have
possible applications, e.g. in the case of intracellular transport,
which typically takes place on one-dimensional tracks. In case of
uni-directional stochastic motion imperfections of the tracks
generically lead to local effects, as far as physiological relevant
particle concentrations are considered. By contrast we have shown that
in case of bi-directional motion condensation of particles is generically
observed in the presence of strong disorder, including the possibility
of an effective control of the particle velocity according to Eq.~(\ref{z_st}).

LS and RJ acknowledge support by the Deutsche Forschungsgemeinschaft
under Grant No. SA864/2-1. This work has been supported by a
German-Hungarian exchange program (DAAD-M\"OB), by the Hungarian
National Research Fund under grant No OTKA TO34138, TO37323, MO45596
and M36803.\\


\begin{thebibliography}{}
\bibitem{note}
        Some results for the random $XX$-chain can be transformed from the
        solution of the RG equations for the random transverse-field Ising
        spin chain, see in Ref \cite{DF,i02}.
\bibitem{note2}
At the critical point, where the width of the front is divergent the
two large barriers are not independent, which could explain
the logarithmic correction in Fig.\ref{fig:critical}b.
\end{thebibliography}
\end{document}